\documentstyle[12pt,graphicx]{article}
\newcommand{\n}{{\not\hspace{-0.5ex}\nabla}}
\newcommand{\A}{{\not\hspace{-0.5ex}A}}
\newcommand{\N}{{\not\hspace{-0.8ex}D_L}}
\newcommand{\D}{{\not\hspace{-0.8ex}D}}
\begin{document}
\title{Vacuum Polarization of a Charged Massless Fermionic Field by a
Magnetic Flux in the Cosmic String Spacetime}
\author{J. Spinelly$^{1,2}$ {\thanks{E-mail: spinelly@fisica.ufpb.br}}  
and E. R. Bezerra de Mello$^{2}$ \thanks{E-mail: emello@fisica.ufpb.br}\\
1.Departamento de F\'{\i}sica-CCT\\
Universidade Estadual da Para\'{\i}ba\\
Juv\^encio Arruda S/N, C. Grande, PB\\
2.Departamento de F\'{\i}sica-CCEN\\
Universidade Federal da Para\'{\i}ba\\
58.059-970, J. Pessoa, PB\\
C. Postal 5.008\\
Brazil}
\maketitle       
\begin{abstract}
We calculate the vacuum averages of the energy-momentum tensor associated with a
massless left-handed spinor fields due to magnetic fluxes on idealized cosmic 
string spacetime. In this analysis three distinct configurations of magnetic fields 
are considered: {\it{i)}} a homogeneous field inside the tube, {\it{ii)}} a magnetic 
field proportional to $1/r$  and {\it{iii)}} a cylindrical shell with 
$\delta$-function. In these three cases the axis of the infinitely long tubes of 
radius $R$ coincides with the cosmic string. In order to proceed with these 
calculations we explicitly obtain the Euclidean Feynman propagators associated with 
these physical systems. As we shall see, these propagators possess two distinct 
parts. The first are the standard ones, i.e., corresponding to the spinor Green 
functions associated with the massless fermionic fields on the idealized cosmic 
string spacetime with a magnetic flux running through the line singularity. The 
second parts are new, they are due to the finite thickness of the radius of the 
tubes. As we shall see these extra parts provide relevant contributions to the 
vacuum averages of the energy-momentum tensor.  
\vspace{1pc}
\end{abstract}

\maketitle

\section{Introduction}

Different types of topological defects may have been formed during the phase
transition in the early universe \cite{Vilenkin}. Depending on the topology of
the vacuum manifold they are domain walls, strings, monopoles and textures. Cosmic
strings have gained some interest recently since they are considered as 
a good candidate to explain some components of anisotropy on the cosmic microwaves
background \cite{Sarangi}, gamma ray bursts \cite{Berezinski}, gravitational waves 
\cite{Damour} and highest energy cosmic rays \cite{Bhattacharjee}. Also they are
thought to be important for the structure formation in the universe due to their
huge energy density per unit length \cite{Kible}. \footnote{ For cosmic strings 
formed at grand unified theory, the energy densities is of order than $10^{21}Kg/m$ 
and their radius $10^{-32}m$.}

A classical field theory model which presents stringlike solutions is the Abelian
Higgs model \cite{Nielsen}. These solutions, also called ``vortices'', correspond to
infinitely long objects. They have a core radius proportional to the inverse of
the Higgs mass and magnetic flux tube with radius proportional to the inverse of
the gauge boson mass.

Coupling the Abelian Higgs model with gravity, Garfinkle \cite{Garfinkle} showed 
that, as in flat spacetime, there exist static cylindrically symmetric solutions 
representing vortices. He also shown that asymptotically the spacetime around the 
cosmic string is a Minkowiski one minus a wedge. Its core has a non-zero thickness, 
and the magnetic field vanishes outside of it. A complete analysis about the 
structure of the Higgs and magnetic fields near the $U(1)-$gauge cosmic strings can 
only be provided numerically.

The vacuum polarization effects due to a magnetic field confined in a tube 
of finite radius in Minkowiski spacetime were first analysed by Serebryanyi 
\cite{Serebryani}. A few years later, Guimar\~aes and Linet \cite{Emilia} and 
Linet \cite{Linet}, calculated these effects for a charged massless scalar and
fermionic fields, respectively, on a idealized cosmic string spacetime. There a 
magnetic field running through the line singularity was considered. As a 
consequence, the renormalized vacuum expectation value associated with the
energy-momentum tensor, $\langle T_{\mu\nu}(x)\rangle_{Ren.}$, presents contributions
coming from the geometry of the spacetime and also the magnetic flux. In order 
to develop these calculations the respective Green functions were obtained. More 
recently Sriramkumar \cite{Sri} has calculated the vacuum fluctuations of current 
and energy densities for a massless charged scalar field around an idealized cosmic 
string carrying a magnetic flux. There the Green function was obtained taking into 
account the presence of the vector potential in the differential operator, 
$D_\mu=\partial_\mu-ieA_\mu$, which presents the advantage of calculating the 
two-point function without imposing any boundary condition on the field 
\footnote{Absorbing the gauge field on the definition of the matter fields implies 
an overall phase shift in the Green function. However this procedure provides to 
$\langle T_{\mu\nu}(x)\rangle_{Ren.}$ the same result as obtained considering 
explicitly the presence of the gauge field in the differential operator.}.  

The analysis of the vacuum polarization effects on a massless charged 
scalar field by a magnetic flux confined in a cylindrical tube of finite radius 
in a cosmic string background was performed in \cite{Spinelly}. There three 
distinct configurations of magnetic field specified below have been considered.
In all of them the axis of the infinitely long tube of radius $R$ 
coincides with the cosmic string. Calculating the renormalized vacuum expectation 
value (VEV) of the square of the field, $\langle\Phi^*(x)\Phi(x)\rangle_{Ren}$,  
and the energy-momentum tensor, $\langle T^\nu_\mu(x)\rangle_{Ren}$, it was 
observed that these quantities present two contributions for each model of magnetic 
flux. The first are the standard ones due to the conical geometry of the spacetime 
and the magnetic flux. The second contributions are corrections due to the finite 
thickness of the radius of the tube. These extra terms provided relevant 
contributions for points outside the tube. Specifically, for the third model this 
contribution is a long-range effect, i.e., it is as relevant as the standard one up 
to a distance which exceed the radius  of the observable universe. 

In this paper we return to this problem and extend the above analysis to the
charged massless fermionic fields. As we shall see, the renormalized VEV of the 
energy-momentum tensors also present two distinct contributions. Moreover due mainly 
to interaction of the spin degrees of freedom of the fermionic fields with the 
magnetic fluxes, the corrections due to the nonzero thickness of the radius of the 
tube are composed by two terms. Unfortunately it is not possible to provide 
analytical informations about the radial behavior of these corrections, and only 
numerical analysis can do that. Also, additional couplings between the spin with 
the geometry are also present in this system.

The idealized model for a infinitely long straight static cosmic string 
spacetime can be given in cylindrical coordinates by the line element:
\begin{equation}
ds^2=-dt^2+dr^2+\alpha^2 r^2d\theta^2+dz^2 \ , 
\label{1}
\end{equation}
where $\alpha=1-4\mu$ is a parameter smaller than unity which codifies the 
presence of a conical two-surface $\left( r,\theta \right)$. In fact for a 
typical Grand Unified Theory, $\alpha=1-O(10^{-6})$.

We shall consider the presence of a magnetic field along the $z$-direction 
assuming that the field has a finite range in the radial coordinate. We are
particularly interested in the three models:\\ 
i) $H(r)=\frac{\phi}{\alpha\pi R^2}\Theta(R-r)$, homogeneous field inside;\\ 
ii) $H(r)=\frac{\phi}{2\pi\alpha Rr}\Theta(R-r)$, field proportional to $1/r$;\\ 
iii) $H(r)=-\frac{\phi}{2\pi\alpha R}\delta(r-R)$, cylindrical shell,\\ 
where $R$ is the radial extent of the tube, and $\phi$ is the total flux. The ratio of the flux to the quantum flux $\phi_{o}$, can be expressed by 
$\phi/\phi_0=N+\gamma$, where $N$ is the integer part and $0<\gamma<1.$\footnote{In this paper we are considering $\hbar=G=c=1$.}

Although the structure of the magnetic field produced by a $U(1)-$gauge cosmic
string cannot be presented by any analytical function, its influence on the
vacuum polarization effects of charged matter fields takes place for sure. So
in this case, the geometric and magnetic interactions provide contributions. The
relevant physical question is how important they are. In this paper, together
with previous ones, by Guimar\~aes and Linet, Linet himself and us, we try to
answer that question. In our analysis we added an extra ingredient on this 
investigation not yet considered.

As we have already mentioned, considering a finite thickness for the magnetic flux,
additional contributions to the vacuum averages are obtained. Because there is no 
possibility to analysis these effects for a realistic gauge cosmic string, 
we consider idealized configurations to the magnetic field. This allows us to 
develop an analytical procedure, and as we shall see, even for these configurations,
some general informations regarding to the influence of the nonvanishing thickness 
of the magnetic flux on the vacuum polarization can be extracting.

This paper is organized as follows. In section $2$, we explicitly construct the
spinor Feynman propagators for three different models of magnetic fields. Having 
these Green functions, in section $3$ we calculate the formal expression to the respective renormalized 
vacuum expectation values of the energy-momentum tensor 
$\langle T_\mu^\nu(x)\rangle_{Ren.}$ associated with the left-handed fermionic 
fields. Because non standard contributions to these VEV do not provide concrete 
informations about their radial behavior, in section $4$ we present our numerical 
analysis about them for specific values for the parameters $\alpha$,
$\gamma$ and $N$. Finally we leave for section $5$ our concluding remarks.  

\section{Spinor Feynman Propagator}

The Feynman propagator associated with a charged fermionic field, $S_F(x,x')$, 
obeys the following differential equation:
\begin{equation}
\left(i\n +e\A -M\right)S_F(x,x')=\frac1{\sqrt{-g}}
\delta^4(x,x')I_4 \ ,
\end{equation}
where $g=det(-g_{\mu\nu})$. The covariant derivative operator reads
\begin{equation}
\n=e^\mu_{(a)}\gamma^{(a)}(\partial_\mu+\Gamma_\mu) \ ,
\end{equation}
$e^\mu_{(a)}$ being the vierbein satisfying the condition $e^\mu_{(a)}e^\nu_{(b)}
\eta^{(a)(b)}=g^{\mu\nu}$ and $\Gamma_\mu$ is the spin connection given in terms
of flat spacetime $\gamma$ matrices by
\begin{equation}
\Gamma_\mu=-\frac14\gamma^{(a)}\gamma^{(b)}e^\nu_{(a)}e_{(b)\nu;\mu} \ ,
\end{equation}
and 
\begin{equation}
\A=e^\mu_{(a)}\gamma^{(a)}A_\mu \ .
\end{equation}

In the Appendix we show that if a bispinor $D_F(x,x')$ satisfies the differential 
equation
\begin{eqnarray}
\label{D}
\left[-{\cal{D}}^2+\frac14{\cal R}+ieg^{\mu\nu}(D_\mu A\nu)-ie\Sigma^{\mu\nu}F_{\mu\nu}+    
2ieg^{\mu\nu}A_\nu\nabla_\nu\right.
\nonumber\\
\left.+e^2g^{\mu\nu}A_\mu A_\nu+M^2\right]D_F(x,x')=-\frac1{\sqrt{-g}}\delta^4(x,x')I_4 \ ,
\end{eqnarray}
with
\begin{equation}
\Sigma^{\mu\nu}=\frac14[\gamma^\mu,\gamma^\nu] \ ,
\end{equation}
$\cal{R}$ being the scalar curvature and the generalized d'Alembertian operator 
given by
\begin{eqnarray}
{\cal{D}}^2=g^{\mu\nu}\nabla_\mu\nabla_\nu=g^{\mu\nu}\left(\partial_\mu\nabla_\nu
+\Gamma_\mu\nabla_\nu-\Gamma^\alpha_{\mu\nu}\partial_\alpha\right) \ ,\nonumber
\end{eqnarray}
then the spinor Feynman propagator may be written as
\begin{equation} 
S_F(x,x')=\left(i\n+e\A+M\right)D_F(x,x') \ .
\end{equation}

Now after this brief introduction about the calculation of spinor Feynman propagator,
let us specialize it to the cosmic string spacetime in the presence
of a magnetic field along the $z-$direction. We shall choose the following
base tetrad:
\begin{equation}
e^\mu_{(a)} = \left( \begin{array}{cccc}
  1& 0 & 0 & 0 \\
  0 & \cos\theta & -\sin\theta/\alpha r 
 & 0\\ 
0 & \sin\theta & \cos\theta/\alpha r
 & 0\\ 
0 & 0 & 0& 1
                      \end{array}
               \right) \ .
\end{equation}

In order to provide an explicit expression to the differential operators above,
we shall adopt the same representation as in \cite{BD} to the $\gamma-$matrices
in Minkowiski spacetime.

As to the four vector potential we have
\begin{equation}
A_{\mu}=(0,0,A(r),0) \ ,
\end{equation} 
with
\begin{equation}
A(r)=\frac{\phi}{2\pi}a(r) \ . 
\end{equation}

For the two first models, we can represent the radial function $a(r)$ by:
\begin{equation}
\label{a1}
a(r)=f(r)\Theta (R-r)+\Theta (r-R) \ ,
\end{equation}
with
\begin{eqnarray}
\label{a2}
f(r)=\left\{\begin{array}{cc}
r^2/R^2,&\mbox{for the model ({\it{i}}) and}\\
r/R,&\mbox{for the model ({\it{ii}}).}
\end{array}
\right.
\end{eqnarray}
For the third model,
\begin{equation}
\label{a3}
a(r)=\Theta(R-r).
\end{equation}
In this spacetime the only non-zero spin connection is
\begin{equation}
\Gamma_2=\frac i2(1-\alpha)\Sigma^3
\end{equation}
and for the Christofell symbols we have:
\begin{equation}
\Gamma_{22}^1=-\alpha^2 r \ , \ \ \Gamma_{12}^2=\Gamma_{21}^1=1/r \ .
\end{equation}

Defining by ${\cal{K}}(x)$ the $4\times 4$ matrix differential operator which acts
on $D_F(x,x')$ in (\ref{D}), for this physical system we obtain:
\begin{eqnarray}
{\cal{K}}(x)&=&-{\Delta}-\frac i{\alpha^2 r^2}(1-\alpha)\Sigma^3\partial_\theta
-eH(r)\Sigma^3+\frac1{4\alpha^2 r^2}(1-\alpha)^2\nonumber\\
&-&\frac e{\alpha^2 r^2}(1-\alpha)A(r)\Sigma^3+\frac{2ie}{\alpha^2 r^2}A(r)\partial_\theta+\frac{e^2}{\alpha^2 r^2}A^2(r)+M^2\ , 
\end{eqnarray}
where
\begin{equation}
\Sigma^3=\left( \begin{array}{cccc}
  \sigma^3& 0\\ 
0 & \sigma^3
                      \end{array}
               \right) \ ,
\end{equation}
and 
\begin{equation}
{{\Delta}}=-\partial_t^2+\partial_r^2+\frac 1r\partial_r+\frac 1{\alpha^2 r^2}
\partial^2_\theta+\partial_z^2 \ .
\end{equation}

We can see that the differential operator above explicitly  exhibits, besides the
ordinary d'Alembertian operator on the conical spacetime, four different types of
interaction terms: $(i)$ the usual charge-magnetic field, $(ii)$ the 
spin-magnetic field, $(iii)$ the spin-geometry and $(iv)$ the spin-charge-geometry.
All of the last three interactions were absent in the analogous differential 
operator used to define the scalar Green function in \cite{Spinelly}.

Moreover, as we can see the operator ${\cal{K}}(x)$ is diagonal in $2\times2$ blocks.
This means that the two upper components of the Dirac spinor interact with the 
gravitational and magnetic fields in similar way as the two lower components and they
do not interact among themselves.

The system that we want to study consists of massless charged fermionic field in the
cosmic string spacetime in the presence of an Abelian magnetic field along the
$z-$direction. Let us chose a left-handed field. In this case the Dirac equation
and the equation which defines the spinor Feynman propagator reduce themselves to
a $2\times 2$ matrix differential equation:
\begin{equation}
\N\chi=0 \ ,
\end{equation}
where
\begin{eqnarray}
\N=i\left[\partial_t-\sigma^{(r)}\left(\partial_r-\frac{(\alpha^{-1}-1)}{2r}\right)-\frac1{\alpha r}\sigma^{(\theta)}\left(\partial_\theta-iA(r)\right)-\sigma^{(z)}
\partial_z\right] \ ,
\end{eqnarray}
with $\sigma^{(r)}={\vec\sigma}.{\hat{r}}$, $\sigma^{(\theta)}={\vec{\sigma}}.
{\hat{\theta}}$ and $\sigma^{(z)}={\vec\sigma}.{\hat{z}}.$

The Feynman two-components propagator obeys the equation
\begin{equation}
i\N S_F^L(x,x')=\frac1{\sqrt{-g}}\delta^4(x,x')I_2 \ ,
\end{equation}
and can be given by
\begin{equation}
S_F^L(x,x')=i\N G^L(x,x') \ ,
\end{equation}
where now $G^L(x,x')$ obeys the $2\times 2$ matrix differential equation:
\begin{equation}
\label{K}
{\bar{K}}(x)G^L(x,x')=-\frac1{\sqrt{-g}}\delta^4(x,x')I_2 \ ,
\end{equation}
with
\begin{eqnarray}
{\bar{K}}(x)&=&-{\Delta}-\frac i{\alpha^2 r^2}(1-\alpha)\sigma^3\partial_\theta
-eH(r)\sigma^3+\frac1{4\alpha^2 r^2}(1-\alpha)^2\nonumber\\
&-&\frac e{\alpha^2 r^2}(1-\alpha)A(r)\sigma^3+\frac{2ie}{\alpha^2 r^2}A(r)\partial_\theta+\frac{e^2}{\alpha^2 r^2}A^2(r)\ .
\end{eqnarray} 

Because of the peculiar diagonal form of the above operator, let us take for $G^L$
the following expression:
\begin{equation}
\label{GL} 
G^L(x,x') = \left( \begin{array}{cccc}
  G_+(x,x')& 0 \\ 
0 & G_-(x,x')
                      \end{array}
               \right) \ .
\end{equation}
So the differential equation (\ref{K}) reduces itself to the two independent ones:
\begin{eqnarray}
\label{G1}
\left[{\Delta}\pm\frac i{\alpha^2 r^2}(1-\alpha)\partial_\theta-\frac{(1-\alpha)^2}
{4\alpha^2 r^2}\pm eH(r)-\frac{2ie}{\alpha^2 r^2}A(r)\partial_\theta \right.\nonumber\\
\left. \pm \frac{e}{\alpha^2 r^2}(1-\alpha)A(r)-\frac{e^2}{\alpha^2 r^2}A^2\right]G_{\pm}(x,x')=\frac1{\sqrt{-g}}\delta^4(x,x') \ .
\end{eqnarray}

Due to the cylindrical symmetry of this system each component of the Euclidean 
Green function can be expressed by
\begin{eqnarray}
G_{\pm}(x,x')=\frac1{(2\pi)^3}\sum_{n=-\infty}^\infty e^{in(\theta-\theta')}
\int_{-\infty}^\infty dk \int_{-\infty}^{\infty} d\omega e^{ik(z-z')}e^{i\omega (\tau-\tau')} g_n^{\pm}(r,r') \ .
\label{11}
\end{eqnarray}

Before to specialize on the specific model, let us write down the non-homogeneous
differential equation obeyed by the unknown function $g_n^\pm(r,r')$.
Substituting (\ref{11}) into (\ref{G1}) and using the standard representation to 
the delta function in the temporal, angular and $z$-coordinates, we arrive at the 
following differential equation for the unknown function $g_n^{\pm}(r,r')$:
\begin{eqnarray}
\left[\frac{d^2}{dr^2}+\frac{1}{r}\frac d{dr}-\frac1{\alpha^2r^2}\left[n^2\pm n(1
-\alpha \mp 2neA)+\frac{(1-\alpha^2)}4\mp e(1-\alpha)A(r)\right. \right. \nonumber\\
 \left.\left.+e^2A^2\right] -\beta^2 \pm H(r) \right]g_n^\pm(r,r')=\frac1{\alpha r}\delta(r-r') \ , \nonumber\\  
\label{12}
\end{eqnarray}
where $\beta^2=k^2+\omega^2$.

It is of our main interest to investigate the vacuum polarization effect for 
external points to the magnetic flux. So, we shall consider solutions of (\ref{12}) 
with both $r$ and $r'$ greater than $R$.

Let us define by $g_n^<(r,r')$ the solution of (\ref{12}) regular at $r\to 0$, and
by $g^>_n(r,r')$ the solution that vanishes at infinity. These two solutions must
satisfy continuity condition at $r=r'$, with their first derivative discontinuous
at this point. Integrating out in the region $r<R$ the inner solutions, corresponding
to $r<r'$, for the first two models we have:
\begin{equation}
g^{\pm <}_n(r,r')=A_{(i)}^\pm H_{i}^\pm(r)\ ,
\label{13}
\end{equation}
for $r<R$ and
\begin{eqnarray}
g^{\pm <}_n(r,r')=B_{(i)}^\pm\left[I_{|\nu_{\pm}|}(\beta r)+E_{(i)}^\pm(\beta R)K_{|\nu_{\pm}|}(\beta r)\right],
\label{14}
\end{eqnarray}
for $R<r<r^{'}$, where 
\begin{equation}
\nu_{\pm}=\frac{(n \pm \frac{1-\alpha}2-\delta)}\alpha \ ,
\end{equation}
with $\delta=\frac{e\phi}{2\pi}=N+\gamma$. $H_{i}^\pm(r)$, for $i=1,2$, represents 
the solution associated with the two first models:
\begin{equation}
H_{1}^\pm(r)=\frac{1}{r}M_{\sigma_{1(\pm)},|\lambda_{1(\pm)}|}\left( \frac{\delta}{\alpha R^2} 
r^2\right),
\label{15}
\end{equation}
and
\begin{equation}
H_{2}^\pm(r)=\frac{1}{\sqrt{r}}M_{\sigma_{2(\pm)},|\lambda_{2(\pm)}|}\left( \zeta r\right),
\label{16}
\end{equation}
with
\begin{equation} 
\sigma_{1(\pm)}=\frac{n}{2\alpha}-\frac{\beta^{2}R^{2}\alpha}{4\delta}
\pm\frac{\alpha+1}{4\alpha} \ ,
\end{equation}
\begin{equation}
\lambda_{1(\pm)}=\frac{n}{2\alpha}\pm\frac{1-\alpha}{4\alpha} \ ,
\end{equation}
\begin{equation}
\sigma_{2(\pm)}=\frac{\delta(2n\pm 1)}{2\alpha(\delta^{2}+\beta^{2}
\alpha^{2}R^{2})^{1/2}}
\end{equation}
and 
\begin{equation}
\lambda_{2(\pm)}=\frac{n}\alpha\pm\frac{1-\alpha}{2\alpha} \ .
\end{equation}
For both functions $H_2^\pm$, $\zeta=\frac{2}{R\alpha}(\delta^2+\beta^2 R^2 
\alpha^2)^{1/2}$. 

In both cases $M_{\sigma,\lambda}$  represents the Whittaker function. The constant 
$E_{(i)}^\pm(\beta R)$ is determined by matching $g_{n}^{\pm<}$ 
and its first derivative at $r=R$. So we get 
\begin{equation}
\label{E1}
E_{(i)}^\pm(\beta R)=\frac{H_i'^\pm(R)I_{|\nu_\pm|}(\beta R)-H_i^\pm(R)I'_{|\nu_\pm|}(\beta R)}{H_{i}^\pm(R)K'_{|\nu_\pm|}(\beta R)-H_i'^\pm(R)K_{|\nu_\pm|}(\beta R)} \ .
\label{17}
\end{equation} 
In all the expressions, $I_\nu$ and $K_\nu$ are the modified Bessel 
functions.

The outer solution of (\ref{12}) is given by
\begin{equation}
g^{\pm>}_{n(i)}(r,r')=D_{(i)}^\pm K_{|\nu_\pm|}(\beta r), 
\quad \hbox{for $r>r'$.}
\label{18} 
\end{equation}

Now, imposing the boundary conditions on $g^{\pm<}_n$ and $g^{\pm>}_n$ at $r=r'$,
we get the following result:
\begin{eqnarray}
\label{gn}
g_n^\pm(r,r')=-\frac1\alpha\left[I_{|\nu\pm|}(\beta r^<)+E^{\pm}_{(i)}(\beta R)K_{|\nu\pm|}(\beta r^<)\right]K_{|\nu\pm|}(\beta r^>) \ .
\label{19}
\end{eqnarray} 
In the above equation $r^>$ ($r^<$) is the larger (smaller) value between $r$ and
$r'$. Substituting (\ref{gn}) into (\ref{11}), developing the sum in $n$ and 
integrating over $k$ and $\omega$ for the first part, the Euclidean Green functions 
acquire the following expression:
\begin{eqnarray}
\label{Ga}
G_\pm(x,x')&=&-\frac{e^{i N\Delta\theta}}{8\pi^2\alpha rr'\sinh u_0}\frac{e^{\mp i\Delta \theta}\sinh(\gamma^\pm u_0/\alpha)+\sinh[(1-\gamma^\pm)u_0/
\alpha]}{\cosh(u_0/\alpha)-\cos\Delta\theta}
\nonumber\\
&-&\frac{1}{4 \pi^{2}\alpha}\int_0^{\infty} d\beta \beta J_{0}\left(\beta\sqrt{(\Delta 
\tau)^{2}+(\Delta z)^{2}}\right)\nonumber\\ 
&&\sum_ne^{i n\Delta\theta}E^\pm_{(i)}(\beta R)K_{|\nu_\pm|}(\beta r)K_{|\nu_\pm|}
(\beta r') \ ,\nonumber\\
\label{19}
\end{eqnarray}
with $\gamma^\pm=(1-\alpha)/2 \mp \gamma$ and
\begin{equation}
\cosh u_{o}=\frac{r^2+{r^{'}}^{2}+(\Delta \tau)^{2}+(\Delta z)^{2}}{2rr^{'}} \ .
\label{20}
\end{equation}

As we can observe, the first term in these Green functions depends only on the
conicity parameter and the fractional part of $\phi/\phi_0$. Moreover, the second
part contains information about the radius of the magnetic tube through the 
constants $E_{(i)}^\pm$.

As to the third model, the solution to $g_{n}^\pm(r,r')$ is:
\begin{equation}
g^{\pm<}_n(r,r')=A^\pm I_{|\nu_\pm|}(\beta r) \ ,
\end{equation}
for $r<R$,
\begin{eqnarray}
g^{\pm<}_n(r,r')&=&B^\pm\left[I_{|\epsilon_\pm|}(\beta r)+E^\pm(\beta R)K_{|\epsilon_\pm|}(\beta r)\right] \ ,
\end{eqnarray}
for $R<r<r^{'}$ and
\begin{equation}
g^{\pm>}_n(r,r')=D^\pm K_{|\epsilon_\pm|}(\beta r), 
\label{22} 
\end{equation}
for $r>r^{'}$. In the above equation $\epsilon_\pm=1/\alpha(n\pm(1-\alpha)/2)$.
Again, the coefficients $A^\pm$, $B^\pm$ and $D^\pm$ can be determined by imposing
boundary conditions at $r=R$ and $r=r'$. However, due to the expression
of the magnetic field in this model is concentrate as a $\delta-$function
at the cylindrical shell, there happens a discontinuity condition in the
first derivative of $g^{\pm<}$ at $r=R$. The rest being the same. So
using these facts we obtain:
\begin{eqnarray}
g^{\pm}_n(r,r')&=&-\frac1\alpha\left[I_{|\epsilon_\pm|}(\beta r^<)+E^\pm(\beta R)K_{|\epsilon_\pm|}(\beta r^<)\right]K_{|\epsilon_\pm|}(\beta r^>) \ ,
\end{eqnarray}
where now
\begin{equation}
E^\pm(\beta R)=\frac{S^\pm(\beta R)}{P^\pm(\beta R)}
\label{E2}
\end{equation}
with
\begin{eqnarray}
S^\pm(\beta R)=I_{|\epsilon_\pm|}(\beta R)I'_{|\nu_\pm|}(\beta R)-I_{|\nu_\pm|}(\beta R)
I^{'}_{|\epsilon_\pm|/\alpha}(\beta R) \pm\frac{\delta}{\alpha R}I_{|\epsilon_\pm|}(\beta R)I_{|\nu_\pm|}(\beta R)&& \nonumber
\end{eqnarray}
and
\begin{eqnarray}
P^\pm(\beta R)=I_{|\nu_\pm|}(\beta R)K^{'}_{\epsilon_\pm}(\beta R)-I_{|\nu_\pm|}^{'}(\beta R)K_{|\epsilon_\pm|}(\beta R)\mp \frac{\delta}{\alpha R} K_{|\epsilon_\pm|}(\beta R)I_{|\nu_\pm|}(\beta R).&& \nonumber
\end{eqnarray}

Finally, substituting the expression found to $g^\pm_n$ above into 
(\ref{11}), and adopting similar procedure as we did in the two previous 
cases, we obtain:
\begin{eqnarray}
\label{Gb}
G_\pm(x,x')=&-&\frac1{8\pi^2 \alpha rr'\sinh u_0}\frac{e^{i\mp\Delta\theta}
\sinh({\bar{\gamma}}u_0/\alpha)+\sinh[(1-{\bar{\gamma}})u_0/\alpha]}
{\cosh(u_{o}/\alpha)-\cos\Delta\theta}
\nonumber\\
&-&\frac1{4 \pi^{2}\alpha}\int_0^{\infty} d\beta \beta J_{0}\left(\beta\sqrt{(\Delta 
\tau)^{2}+(\Delta z)^{2}}\right)\times
\nonumber\\
&&\sum_{n=-\infty}^\infty e^{in\Delta\theta}
E^\pm(\beta R)K_{|\epsilon_\pm|}(\beta r)K_{|\epsilon_\pm|}(\beta r') \ ,
\label{24}
\end{eqnarray}
with ${\bar{\gamma}}=(1-\alpha)/2$.

Here the first term of the Green function depends only on the 
conicity parameter.

Although all expressions to $g^<_n$ and $g^>_n$ present dependence on the 
radial coordinates $r$ and $r'$, their dependence on $r'$ are implicitly contained 
in the coefficients which multiply the functions of $r$. These dependences appear as 
consequence of the boundary conditions obeyed by them at $r=r'$.

\section{Computation of $\langle\hat{T}_{00}\rangle_{Ren}$}

The vacuum expectation value of the energy-momentum tensor associated with the 
system under investigation can be obtained in the book by Grib {\it et al}
\cite{Grib}, combining respectively the classical expressions to this tensor 
associated with the fermionic field in presence of electromagnetic interaction
in a curved spacetime. It is given by:
\begin{eqnarray}
\langle\hat{T}_{\mu\nu}\rangle&=&\frac14\lim_{x'\to x}tr\left[\sigma_\mu(D_\nu-{\bar{D}}_{\nu'}) -\sigma_\nu(D_\mu-{\bar{D}}_{\mu'})\right]S_F^L(x,x') \ , \label{45}
\end{eqnarray}
where $D_\sigma=\nabla_\sigma-ieA_\sigma$, the bar denotes complex 
conjugate and $\sigma^\mu=(I_2,\sigma^{(r)},\sigma^{(\theta)},\sigma^{(z)})$ . 

In order to take into account the presence of the three magnetic field 
configurations, given previously, we write the vector potential in the form 
$A_\mu=(0,0,\frac{\phi a(r)}{2\pi},0)$, with $a(r)$ being given by 
(\ref{a1}) and (\ref{a2}), for the first two cases and by (\ref{a3}) for the third 
case. The spinor Green functions are expressed in terms of the bispinor $G^L(x,x')$ 
given by (\ref{GL}) with $G_\pm(x,x')$ given by (\ref{Ga}) for the two first models 
and (\ref{Gb}) for the third one.

For simplicity let us calculate $\langle T_{00}(x)\rangle$ only. The other 
components of this tensor can be obtained by using the conservation condition
\begin{equation}
\nabla_\mu\langle T^\mu_\nu\rangle_{Ren.}=0 \ ,
\end{equation}
and the vanishing of trace \footnote{In general the trace of the renormalized
VEV of the energy-momentum tensor is equal to $\frac 1{16\pi^2}Tr a_2$ 
\cite{Christensen}. However, because this spacetime is locally flat and there is
no magnetic field outside the tube, $a_2=0.$}
\begin{equation}
\langle T^\mu_\mu\rangle_{Ren.}=0 \ .
\end{equation}  
Fortunately only the second order time derivative provides a nonzero contribution 
to $\langle T_{00}(x)\rangle$. In fact all the other terms go to zero in the
coincidence limit and/or after taking the trace over the Pauli matrices. 
Moreover, because the bispinor depends on the time variable with $t-t'$, we
finally have:
\begin{eqnarray}
\langle T_{00}(x)\rangle&=&-\lim_{x'\to x}tr(\partial_t^2G(x,x'))\nonumber\\
&=&\lim_{x'\to x}tr(\partial_\tau^2G_E(x,x')) \ ,
\end{eqnarray}
where we have made a Wick rotation on the above equation.

However the calculation of the above expression provides a divergent 
result. In order to obtain a finite and well defined expression, we must
apply in this calculation some renormalization procedure. The method which we
shall apply here is the point-splitting renormalization procedure. The basic idea
of this method consists of subtracting from the Green function all the divergences
which appear in the coincidence limit. In \cite{Adler} Adler {\it et al} observed 
that the singular behavior of the Green function in the coincidence limit has the 
same structure as the Hadamard one. Later Wald \cite{Wald} introduced a 
modification to this technique in order to provide the correct result for the
trace anomaly. In this way the renormalized vacuum expectation value of the 
zero-zero component of the energy-momentum tensor can be given by:
\begin{eqnarray}
\langle T_{00}(x)\rangle_{Ren.}&=&\lim_{x'\to x}tr[\partial_\tau^2G_E(x,x')-\partial_\tau^2G_H(x,x')] \ .
\end{eqnarray}
Because this spacetime is locally flat, the Hadamard function coincides with the 
Euclidean Green function in a flat spacetime: 
\begin{equation}
G_H(x,x')=\frac1{4\pi}\frac1{(x-x')^2}I_2 \ .
\end{equation}
As it was already mentioned, (\ref{Ga}) and
(\ref{Gb}) present two distinct contributions; the first ones contain informations
about the geometrical structure of the spacetime and the fractional part of the
magnetic flux, and the second, the corrections due to the nonzero thickness
of the radius of tube. Moreover, in the calculation of the VEV, only their first
contributions are divergent in the coincidence limit, the second ones are finite. 
Finally, explicitly exhibiting these remarks, we write down the renormalized VEV of 
the zero-zero component of the energy-momentum tensor by:
\begin{equation}
\langle T_{00}(x)\rangle_{Ren.}=\langle T_{00}(x)\rangle_{Reg.}+
\langle T_{00}(x)\rangle_{C} \ ,
\end{equation}
where
\begin{eqnarray}
\langle T_{00}(x)\rangle_{Reg.}&=&\lim_{x'\to x}\left[\partial_\tau^2G_+(x,x')
+\partial_\tau^2G_-(x,x')-2\partial_\tau^2G_H(x,x')\right]
\end{eqnarray}
and 
\begin{equation}
\langle T_{00}(x)\rangle_{C}=\lim_{x'\to x}\partial_\tau^2G_C(x,x') \ .
\end{equation}
Here $G_C(x,x')$ represents the corrections due to the second terms in $G_+$ and
$G_-$ for the three models, as shown next. After some intermediate calculations we 
arrive to the following results:\\
\noindent
$i)$ For the two first models,
\begin{eqnarray}
\label{T1} 
\langle T_{00}(x)\rangle_{Ren.}&=&\frac1{5760\pi\alpha^4 r^4}\left[(\alpha^2-
1)(17\alpha^2+7)+120\gamma^2(\alpha^2-2\gamma^2-1)\right]\nonumber\\
&&+\frac1{4\pi^2r^4\alpha}\int_0^\infty dv v^3\sum_{n=-\infty}^\infty\left[
E_{(i)}^+(vR/r)K_{|\nu_+|}^2(v)\right. \nonumber\\
&&\left.+E_{(i)}^-(vR/r)K_{|\nu_-|}^2(v)\right] \ ,
\end{eqnarray}
for $i=1$ and $2$. \\
\noindent
$ii)$ For the third model,
\begin{eqnarray}
\label{T2} 
\langle T_{00}(x)\rangle_{Ren.}&=&\frac1{5760\pi\alpha^4 r^4}(\alpha^2-
1)(17\alpha^2+7)\nonumber\\
&&+\frac1{4\pi^2r^4\alpha}\int_0^\infty dv v^3\sum_{n=-\infty}^\infty\left[E^+(vR/r)
K_{|\epsilon+|}^2(v)\right. \nonumber\\
&&\left. +E^-(vR/r)K_{|\epsilon-|}^2(v)\right] \ .
\end{eqnarray} 
For all the above expressions the coefficients $E^\pm$ were given in (\ref{E1})
and (\ref{E2}). 

The new terms obtained in (\ref{T1}) and (\ref{T2}) are consequence of the finite 
thickness of the magnetic flux and will be present even in the absence of cosmic 
string. Unfortunately it is not possible to evaluate these corrections analytically.
The reason is because the dimensionless variable of integration $v$ appears not
only in the argument of the modified Bessel function, but mainly because it
appears in the order of the Whittaker functions through the factors $E^\pm$.

Before to provide some qualitative information about the second contributions 
of (\ref{T1}) and (\ref{T2}), we would like to make a few comments about our
results: $(i)$ The first contributions to them depend only on the conicity
parameter $\alpha$, and the fractional part of $\phi/\phi_0$, denoted by 
$\gamma$, for the two first models. $(ii)$ The second contributions, the
corrections, vanish in the limit $R\to 0$. They also depend on the integer
part of $\phi/\phi_{0}$, $N$.

Some qualitative informations about the behaviors of the second contributions
can be provided: although for the three models these contributions present an 
overall $1/r^4$ dependence on the radial coordinate, there exist an additional 
dependences in their integrands by the coefficients $E^\pm$. However quantitative 
informations about these new behaviors can only be provided numerically. So we 
leave for the next section this analysis.

\section{Numerical Analysis of Corrections}

In this section we shall exhibit the quantitative behavior for corrections to the
renormalized VEV of the zero-zero components of the energy-momentum tensor for
the system analysed, considering for the three models specific values of the
parameters $\alpha$, $\gamma$ and $N$. These behavior can only be provided by 
numerical analysis.

The numerical method adopted by us to develop this analysis was numerical routine 
of MAPLE to evaluate integrals. Our first general conclusion is about the 
contributions presented by each term in the corrections. Although the integrals in 
the variable $v$ present contributions with opposite sign due to the coefficient 
$E^+$ and $E^-$, they do not cancel each other. This fact is a consequence, mainly, 
to the preferential spatial direction dictated by the magnetic field on the spinor 
degrees of freedom.

Moreover, although the total contributions to the renormalized VEV of the zero-zero 
component of the energy-momentum tensor, $\langle T_{00}\rangle_{Ren.}$, requires 
in principal an infinite sum of terms, in fact this is not really necessary. The 
sums are very well represented by a few terms only. Expressing this quantity as
\begin{equation}
\langle T_{00}\rangle_{Ren.}=\langle T_{00}\rangle_{Reg.}\left(1-
\Delta_{\gamma,\alpha,N}(x)\right) \ ,
\end{equation}
where $x=R/r$, in our graphs we exhibit only the function $\Delta_{\gamma,\alpha,N}$
for the three models, considering $x$ running in the interval $[0.01,0.1]$. We 
also compare these functions for the three models. Although, we expect that this
function goes to zero in the limit as $x\to 0$, or $r\to\infty$, our numerical 
results are not hundred percent trustful for $x$ smaller than $10^{-4}$. In fact
the behavior of the factors $E_{(i)}^\pm(vx)$ for the two first models and
$E^\pm(vx)$ for the third one for small value of $x$ can be obtained by 
developing a series expansion in power of their argument. Our results for them
are:
\begin{equation}
E_{(i)}\simeq \ (vx)^{2|\nu_\pm|} \ ,
\end{equation}
with
\begin{equation}
\nu_\pm=\frac1\alpha\left[n\pm\frac{(1-\alpha)}2-\delta\right] 
\end{equation}
and
\begin{equation}
E^\pm\simeq \ (vx)^{2|\epsilon_\pm|} \ ,
\end{equation}
with
\begin{equation}
\epsilon_\pm=\frac1\alpha\left[n\pm\frac{(1-\alpha)}2\right]
\end{equation}
From these expressions we can infer that the corrections in fact go to zero for 
large values of the radial distance $r$, i.e., small value of $x$. However the 
numerical program provides very irregular diagrams to the integrands of (\ref{T1}) 
and (\ref{T2}) for small values of the dimensionless variable $v$ when $x$ becomes 
smaller than $10^{-4}$, so we can not trust on the evaluations of the integrals for 
these cases.

In Figs. $1(a)$ and $1(b)$, we present the radial dependence of the function 
$\Delta_{\gamma,\alpha,N}$, for $N=1$ and $N=2$, respectively, considering, 
$\alpha=0.99$ and $\gamma=0.02$. From both figures we can see that the shapes of the 
curves are very similar, however the larger contributions are due to the third 
models. By our numerical analysis these functions are bigger than
unity, indicating that the corrections are in fact more relevant than the standard 
results, $\langle T_{00}\rangle_{Reg.}$, up to a distance one thousand times or, 
even more, larger than the radius of the tube. In the Figs. $2(a)$ and $2(b)$ we
exhibit the logarithmic behavior for the same functions for a smaller scale of the
variable $x$. In the latter is possible to infer a power behavior for the
corrections as expected by analytical analysis.

In Fig. $3$, we exhibit the functions $\Delta_{\gamma,\alpha,N}$ for $N=1$ and
$N=2$. From this graph we observe that the correction due to $N=2$ is larger that
for $N=1$. This result is expected, of course. However we know that the values
adopted here to the magnetic fluxes are only few times larger than the quantum flux, 
so increasing this magnetic flux for sufficiently large value of $N$ the corrections 
becomes more and more relevant than the standard values of 
$\langle T_{00}\rangle_{Reg.}$.
  
Finally in Fig. $4$, we exhibit the behavior of $\Delta_{\gamma,\alpha,N}$ varying 
the parameter $\alpha$ for $N=1$. From this graphs we observe a prominent 
increase for this factor for values of $\alpha$ bigger than $0.95$. Moreover we can 
observe too that this effect becomes more evident in the third model of magnetic 
field. A similar graph has been constructed considering $N=2$, however no 
significant changes were found.

\section{Concluding Remarks}

In this paper we have explicitly exhibited the spinor Green functions associated
with a left-handed charged spinor field on a cosmic string spacetime in the presence 
of external magnetic fluxes for three different configurations, all of them confined 
inside a long tube of radius $R$. Having these Green functions, we calculated the
formal expressions to the renormalized VEV of the zero-zero component of the
energy-momentum tensor, $\langle T_{00}(x)\rangle_{Ren}$. In these calculations we 
observe that two independent contributions were obtained. The first contributions 
are the standard ones due to the conical geometry of the spacetime and the fractional
part of $\phi/\phi_0$. They coincide with the expression found by Linet in 
\cite{Linet}. The second contributions are corrections due to the finite thickness 
of the radius of the tube. Unfortunately, it is not possible, by analytical 
analysis, to provide the complete information about the dependence of these parts
with the radial coordinate, and only by numerical analysis we can do this. From 
this analysis we could observe that the corrections are more relevant than the
standard results for distances up to $10^3$ bigger than the radius of the tube.
Our results also suggest that the corrections are almost independent of the specific
form of magnetic field, and that they increase for higher values of the magnitude
of the magnetic field. Analysing the behavior of the corrections, $\Delta_{\gamma,
\alpha,N}$ as function of $\alpha$, we also observe that they become more relevant 
when this parameter becomes closer than unity.

So, from all our numerical analysis we conclude that considering a nonvanishing
radial extension to the magnetic flux in the calculation of the vacuum polarization,
very important contributions take place. In fact these contributions become more
relevant than the standard ones. The main reason for this fact is because no matter 
how big is the magnitude of the magnetic flux, the standard contribution depends 
only on the fractional part of $\phi/\phi_0$.

Although the models investigated here in fact do not correspond to the realistic
configuration associated with a $U(1)-$gauge cosmic string, they provide some
improvement in the calculation of vacuum polarization effects when compared with
the ideal case; moreover our results give some general informations. They suggest
that considering the influence of the nonzero thickness of the radial extent of the
magnetic filed, relevant contributions to the induced energy densities 
associated with massless spinor fields are obtained, and more, these corrections
are almost insensitive to the specific form of the magnetic fields.

Allen {\it et all} \cite{Allen} have investigated the vacuum polarization of a 
massless scalar field on the cosmic string spacetime considering generically the 
effect of the internal structure of the string's core on the metric tensor. In 
this analysis the respective Green function also presents two parts, being one
of them due to the non-zero core radius of the string. A more realistic treatment
of the vacuum polarization effect associated with a charged field by a gauge
cosmic string, must to consider the effect of both non-zero thicknesses for the
magnetic field and the metric tensor. This analysis deserves to be analyzed
appropriately in near future.

{\bf{Acknowledgments}}
\\       \\
J. S. wants to thank the support given by the physics department of Universidade 
Estadual da Para\'{\i}ba (UEPb) and E. R. B. M. to Conselho Nacional de 
Desenvolvimento Cient\'\i fico e Tecnol\'ogico (CNPq.) for partial financial support.

\section{Appendix. Square of the Dirac Operator}

In this appendix we want to find the square of the Dirac operator $i\D=
-\n+ie\A$. The result presents three distinct contributions.\\
$a)$ The gauge field independent term \footnote{This part is very well known. It
is given in the  book by Birrell and Davies \cite{Birrel}}:
\begin{equation}
\gamma^\mu\nabla_\mu\gamma^\nu\nabla_\nu=-{\cal{D}}^2+\frac14{\cal R} \ ,
\end{equation}
being $\cal{R}$ the scalar curvature of the manifold and ${\cal{D}}^2$ the generalized
d'Alembertian operator.\\
$b)$ The term linear in the gauge field has two contributions:
\begin{equation}
-ie\{\gamma^\mu\nabla_\mu A_\nu\gamma^\nu+A_\mu\gamma^\mu\gamma^\nu\nabla_\nu\} \ ,
\end{equation}
which after some intermediate steps can be written as:
\begin{equation}
\label{35}
ie\{g^{\mu\nu}D_\mu A_\nu-\Sigma^{\mu\nu}F_{\mu\nu}+2g^{\mu\nu}A_\nu\nabla_\mu\} \ ,
\end{equation}
where
\begin{equation}
\Sigma^{\mu\nu}=\frac14[\gamma^\mu,\gamma^\nu] \ .
\end{equation}
The second term in (\ref{35}) is a Pauli interaction one. It appears as a 
consequence of the square of the Dirac operator. $F_{\mu\nu}$ is the
usual second-order, antisymmetric field-strength tensor.\\
$c)$ As to the square gauge field term, it is easily obtained being equal to:
\begin{equation}
e^2g^{\mu\nu}A_\mu A_\nu \ .
\end{equation}

In our calculations above we have considered that the $\gamma$ Dirac matrices
obey the anticommutator relation $\{\gamma^\mu,\gamma^\nu\}=-2g^{\mu\nu}$.

\section{Figure Captions}

{\bf Figure 1}: These graphs show the behavior of $\Delta_{\gamma,\alpha,N}$
for the three specific models of magnetic field configurations with:
$(a)$ $N=1$, $\alpha=0.99$ and $\gamma=0.02$. In $(b)$ we change to $N=2$.\hfill\\
[6mm]
{\bf Figure 2}: These graphs present the logarithmic behavior of the previous one
for $N=1$ in smaller scale of the variable $x$.\hfill\\
[6mm]
{\bf Figure 3}: This graph shows the behavior of  $\Delta_{\gamma,\alpha,N}$, for the first model in the case $N=1$ and $N=2$, considering $\alpha=0.99$ and $\gamma=0.02$.\hfill\\[6mm]
{\bf Figure 4}: This diagram presents the behavior of the correction $\Delta_{\gamma,
\alpha,N}$ as function of $\alpha$. It was considered $x=10^{-1}$, $N=1$ and 
$\delta=0.002$.

\begin{figure}[t]
\begin{center}
\includegraphics[width=8cm,angle=-90]{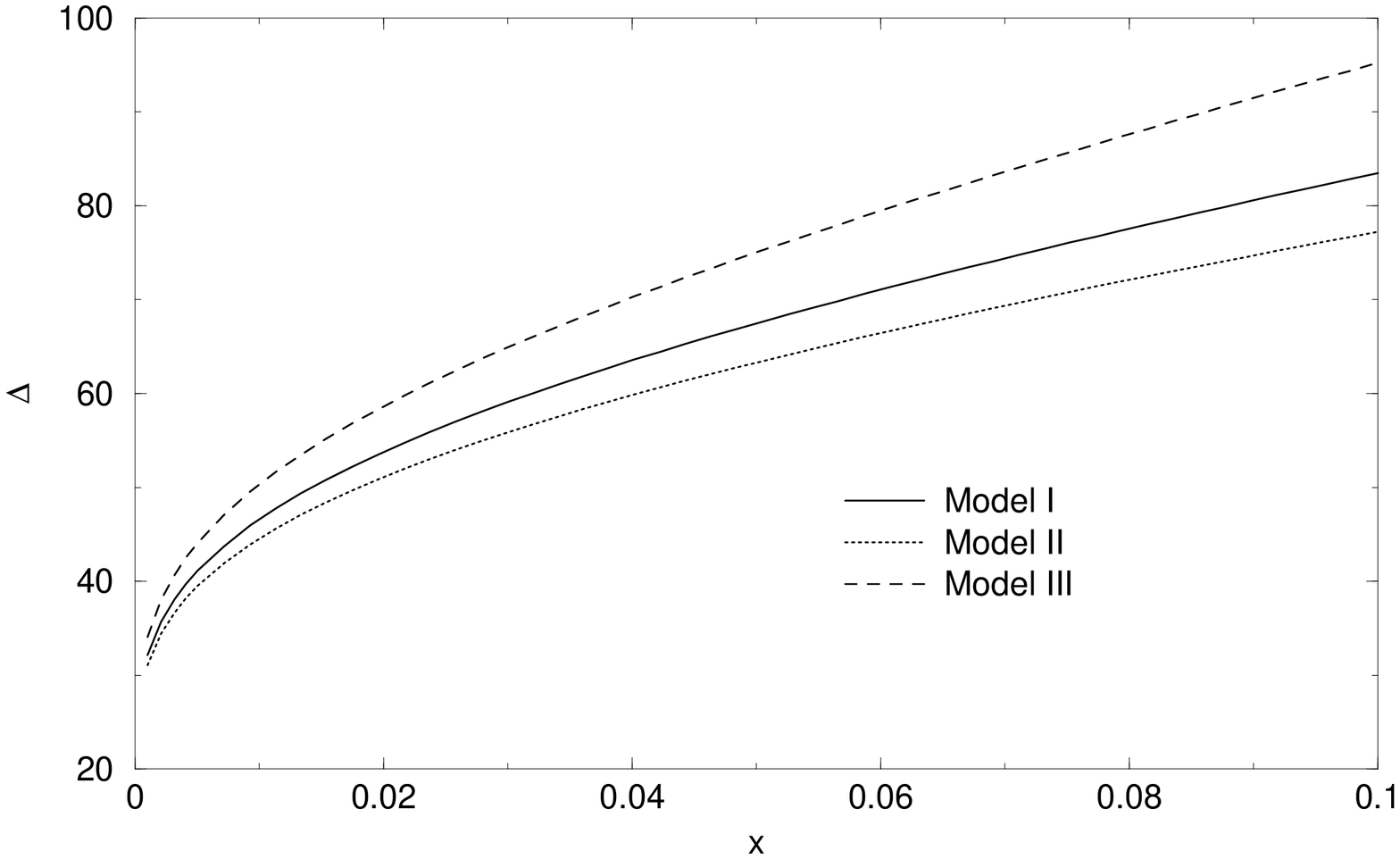}

Fig. 1.a

\includegraphics[width=8cm,angle=-90]{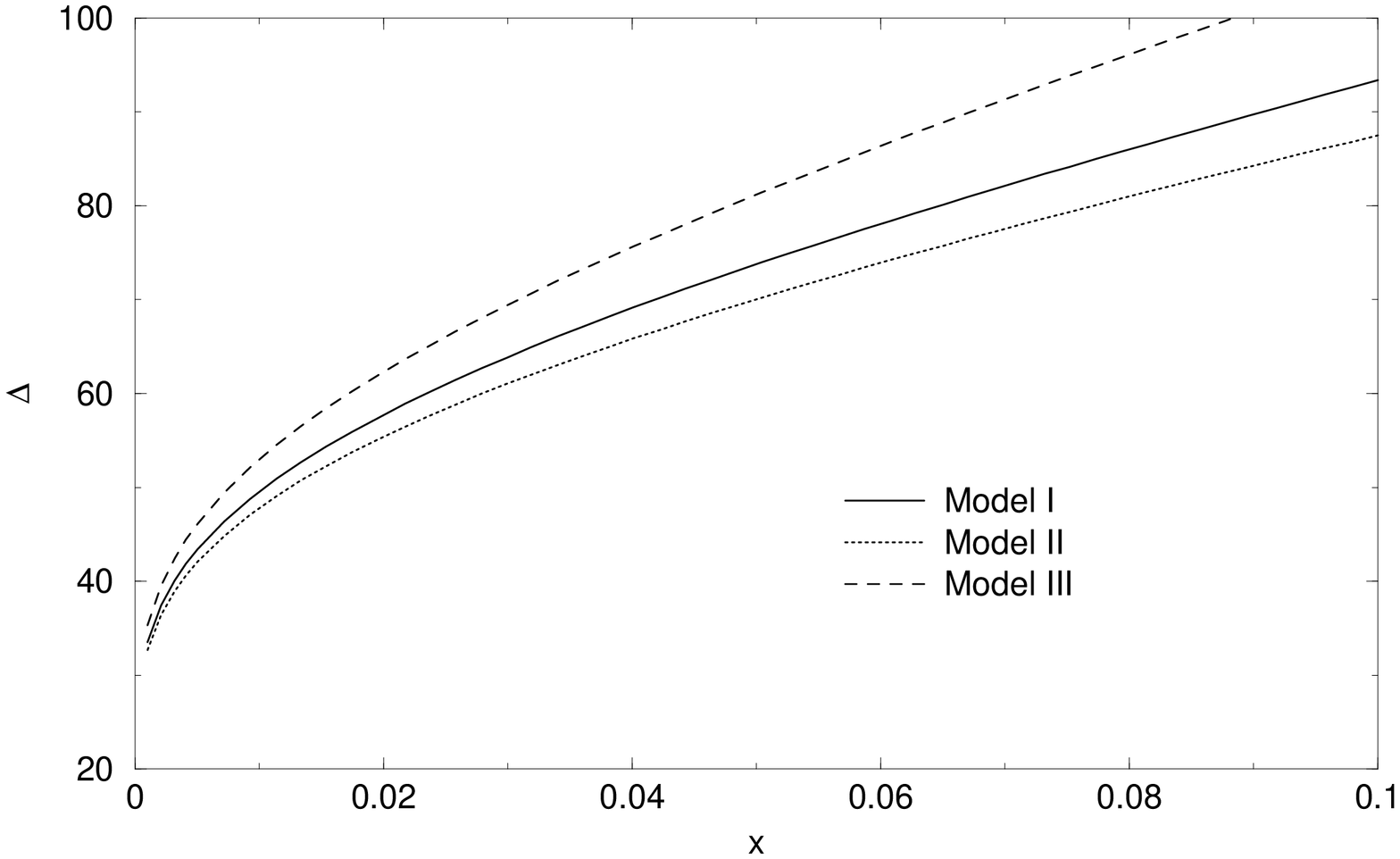}

Fig. 1.b

\end{center}
\end{figure}

\begin{figure}[!]
\begin{center}
\includegraphics[width=8cm,angle=-90]{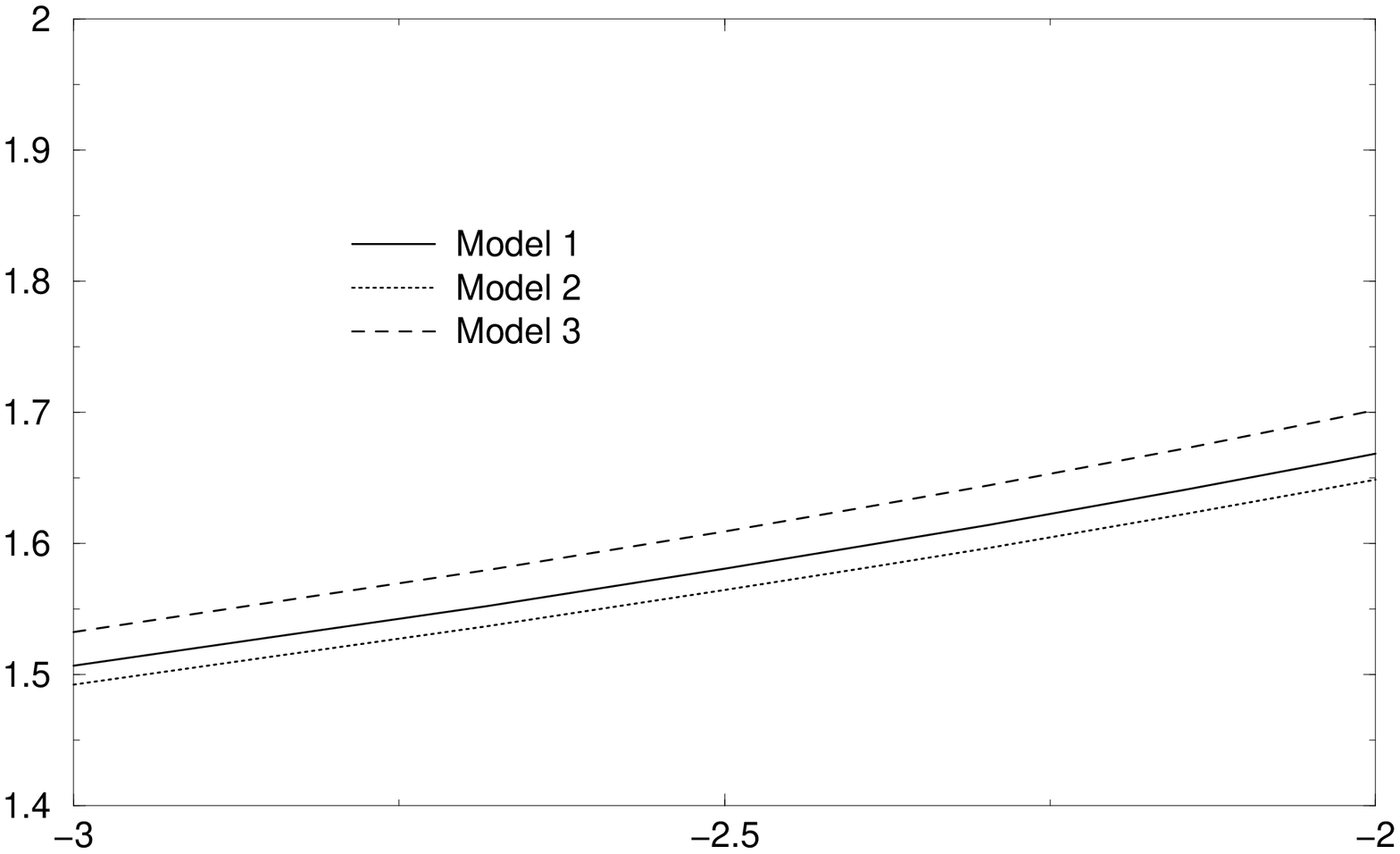}

Fig. 2

\end{center}
\end{figure}

\begin{figure}[!]
\begin{center}
\includegraphics[width=8cm,angle=-90]{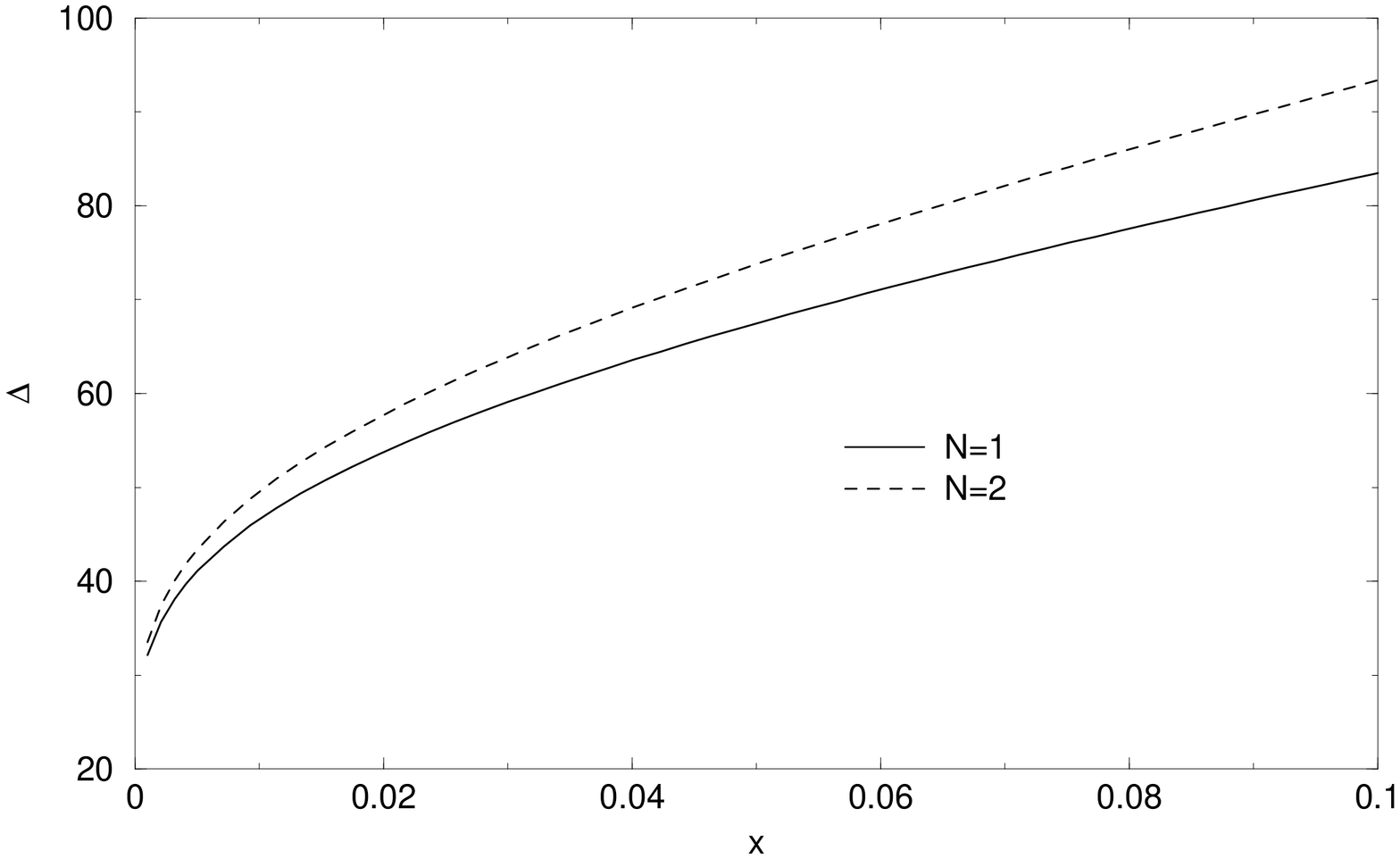}

Fig. 3

\end{center}
\end{figure}

\begin{figure}[!]
\begin{center}
\includegraphics[width=8cm,angle=-90]{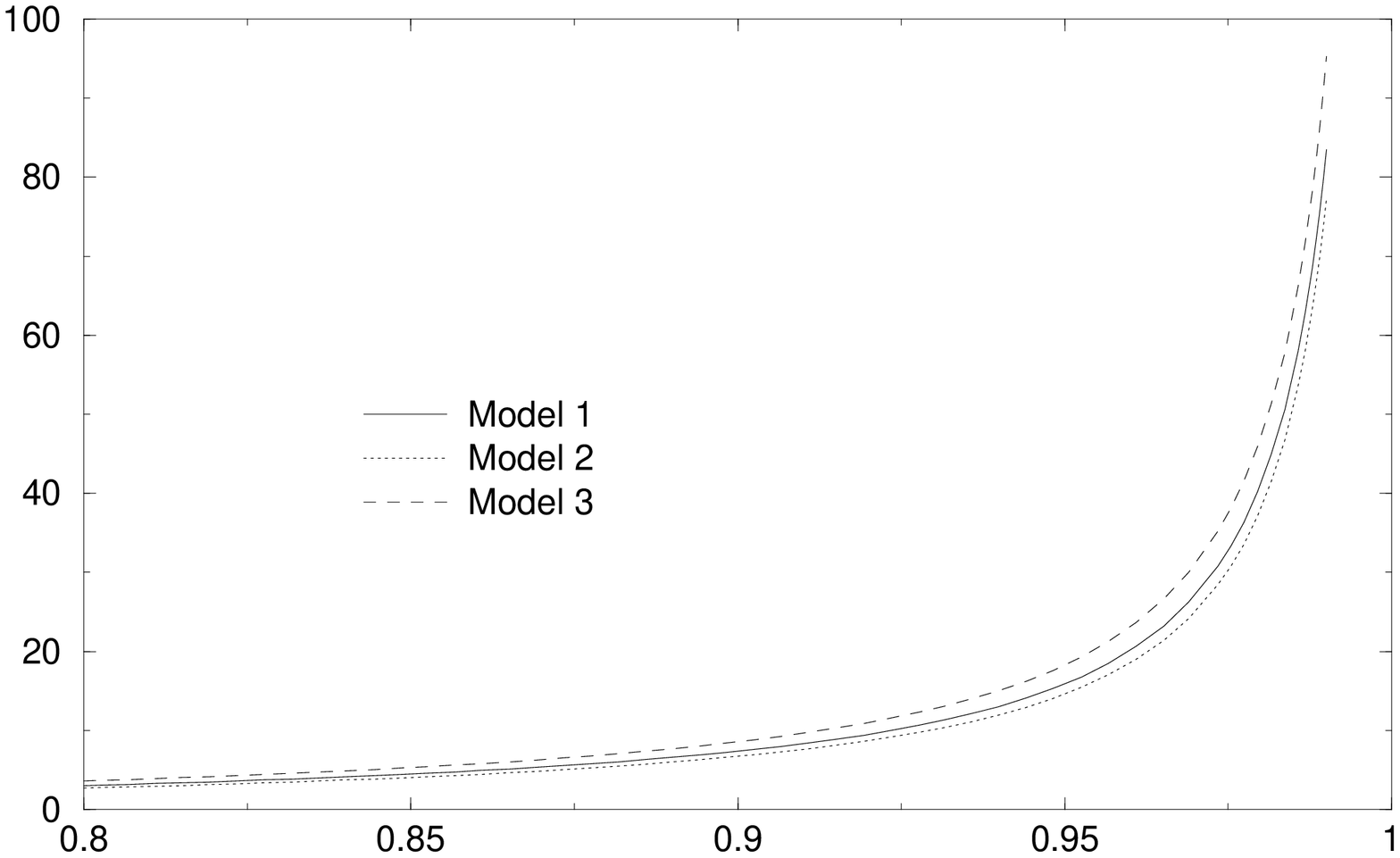}

Fig. 4

\end{center}
\end{figure}

\end{document}